# Nature of Structural Transformations in the $B_2O_3$ Glass under High Pressure

V.V. Brazhkin[1*], Y. Katayama[2], K. Trachenko[3], O.B. Tsiok[1], A.G. Lyapin[1], Emilio Artacho[3], M. Dove[3], G. Ferlat[4], Y. Inamura[2] and H. Saitoh[2]

[1] *Institute for High Pressure Physics RAS, 142190 Troitsk Moscow region, Russia*
[2] *Japan Atomic Energy Agency (JAEA), SPring-8, 1-1-1 Kuoto, Sayo-cho, Sayo-gun, Hyogo, 679-5143, Japan*
[3] *Department of Earth Sciences, University of Cambridge, Downing Street, Cambridge, CB2 3EQ, UK*
[4] *Institut de Mineralogie et Physique des Milieux Condenses,Universite Paris 6, Universite Paris 7, CNRS UMR 7590, IPGP, 140 rue de Lourmel, 75015 Paris, France*

We report the results of the X-ray diffraction study of $B_2O_3$ glass in the pressure interval up to 10 GPa in the 300-700 K temperature range, the results of in-situ volumetric measurements of the glass at pressures up to 9 GPa at room temperature, and first-principles simulations data at pressures up to 250GPa. The behavior of $B_2O_3$ glass under pressure can be described as two broad pressure-overlapping transitions. The first transition starts at P > 1 GPa and proceeds without any changes in thecoordination number of the boron atoms; the other one starts at P > 5 GPa and is accompanied by a gradual increase of coordination number. The second transition is completely reversible; the residual densification of the $B_2O_3$ glass after decompression is associated with the incomplete reversibility of the first transformation. The fraction of boron atoms transferred to the 4-coordination state in the glass at P < 10 GPa is much smaller than was assumed from indirect experimental data [4], but is considerably larger (by tens of times) than was previously suggested by the classical molecular dynamics simulations [5,6]. The observed transformations under both compression and decompression are quite broad, contrary to the assumption in Ref. [7]. On the basis of ab-initio results, we also predict the one more transformation under higher pressures to a super-dense phase, in which B atoms become 5- and 6-fold coordinated.

[*]e-mail address for contacts: brazhkin@hppi.troitsk.ru

Atomic rearrangements and phase transformations in glasses under high pressure and temperature are one of the hottest and most puzzling topics of physics and material science [1-3]. $B_2O_3$ represents an archetypical oxide alongside such glasses as $SiO_2$ and $GeO_2$ [8]. There is a wide range of important boron-oxide-containing optical materials and glass ceramics. Possible minor quantities of $B_2O_3$ in the Earth mantle may strongly affect the dynamics of magmatic melts. It has been thought for about a century that $B_2O_3$ glass structure is almost perfectly suited for a glassy network model, in which the boron atoms are surrounded by three oxygen atoms, whereas oxygen atoms are surrounded by two boron atoms [8]. At the same time, many peculiar features of the $B_2O_3$ glass structure have not been understood to the present day. For example, the question of the fraction of atoms belonging to the $B_3O_6$ boroxol rings remains an open question [6 and refs therein, 9].

The studies of $B_2O_3$ glass under pressure are of considerable importance in understanding glass structure and inter-atomic interactions in glass. However, the most important experimental probes into the glass behaviour under pressure (in-situ X-ray diffraction and volumetric measurements) as well as first-principle simulations of the pressure response have not been reported thus far. X-ray diffraction or the EXAFS studies of $B_2O_3$ glass under pressure are extremely difficult because of the low atomic number of its elements. A neutron diffraction study requires large-sized samples, which restricts the maximum pressure that can be achieved. It is equally difficult to conduct the volumetric measurements of glasses under pressure: most existing techniques, such as direct optical measurements of sizes in transparent high-pressure devices [10] or density measurements from the X-ray absorption data [11], cannot offer sufficiently high accuracy.

The study of the behavior of crystalline $B_2O_3$ under pressure has likewise been fragmentary. The normal pressure phase $B_2O_3$ (I) (space group $P3_1$, $a = 4.336$ Å, $c = 8.340$ Å, $\rho = 2.55$ gr/cm$^3$) [12] undergoes a transition at pressure ~ 4 GPa and high temperature to a high-pressure modification $B_2O_3$ (II) (space group $Ccm2_1$, $a = 4.613$ Å, $b = 7.803$ Å, $c = 4.129$ Å, $\rho = 3.11$ g/cm$^3$) [13], which is stable for at least up to 40 GPa [14]. In $B_2O_3$ (II), boron atoms become four-fold coordinated while oxygens are either two- or three-fold coordinated [13]. The P,T-phase diagram of crystalline $B_2O_3$ was investigated in Refs. [15-17]. The present-day version of the equilibrium phase diagram has been obtained through the use of an in-situ X-ray diffraction study and has been presented in Ref. [18].

As it was found long ago [17,19], the pressure-treated $B_2O_3$ glass shows residual densification $\Delta\rho/\rho$ ~5-10%, depending on the pressure-temperature conditions of treatment. The ex-situ studies of densified glasses by X-ray and neutron diffraction methods point to the breakup of the boroxol rings in the glass structure and to the buckling of the "ribbons" formed by the $BO_3$ triangles, without any significant modification of the short-range order structure and coordination change of the boron and oxygen atoms [20,21]. Some fraction of four-fold coordinated B was observed in the densified $B_2O_3$ glasses obtained by quenching the melt under pressure [22].

The in-situ investigations of $B_2O_3$ glass under pressure have been performed by using Raman and Brillouin spectroscopies [7,23,24], and through inelastic X-ray scattering spectroscopy [4]. Besides, there have been attempts to examine the $B_2O_3$ glass under pressure by molecular dynamics computer simulation, using empirical inter-atomic potentials [5,6]. From the experimental data of Nicholas *et al.* [7] it follows that $B_2O_3$ glass under compression experiences a transformation in the pressure range P~ 6-15 GPa, with the reverse transformation at decompression occurring at P~ 3 GPa. The same authors [7] suggested that this constitutes sharp 1[st] order phase transition, which then in [6] was interpreted as 2[nd] order. This transformation was originally attributed [7] to a change in the coordination of the boron atoms from 3 to 4 in a similar way to crystalline phases. According to Ref. [4], the $B_2O_3$ glass, under compression, features a considerable change in the bonding type in the 6-20 GPa pressure range. This fact was likewise attributed to the change of the short-range order in glass, with the coordination of boron atoms changing from 3 to 4. The molecular dynamics data [5,6] are another testimony to the existence of a smeared transformation in $B_2O_3$ under pressure, accompanied by the modification of the first-neighbor coordination. However, the quantitative estimates of the degree of transformation based on the indirect experiments [4] and those based on the computer simulation data [5] are noticeably different. Thus, according to the estimates made in Ref. [4], the fraction of the boron atoms that had transferred to the 4-coordination state is 50% at P ~ 8 GPa, while the coordination pressure-induced transformation ends at P~ 15 GPa. According to Ref. [5], however, the fraction of the 4-coordinated boron atoms in the glass compressed to 8 GPa is ~1%, and is less than 40% even at 30 GPa. Ref. [6] states that the fraction of the 4-coordinated boron atoms in the glass compressed to 8 GPa is close to zero and it is around 2% at 30 GPa. It remains unclear how well empirical potentials, however sophisticated, can accurately model the non-trivial transformations in $B_2O_3$ glasses under pressure.

Thus, the purpose of this work was to conduct pioneering in-situ structural and volumetric measurements of $B_2O_3$ glasses under pressure, as well as first-principle calculations (see Methods) to elucidate the nature of phase transformations.

The in-situ diffraction structural data for the $B_2O_3$ glass at different pressures are presented in Fig. 1. The structural factor (Fig. 1b) smoothly changes during compression up to 5-6 GPa. On further pressure increase, the structural changes become more pronounced. From the pictures of the total correlation function (Fig. 1c), one can see that at P < 5 - 6 GPa, the $1^{st}$ peak of the total correlation function does not in fact change. The basic changes take place in the distant coordination spheres, beginning from the second one. At P > 5 - 6 GPa, the structural changes begin affecting the 1st coordination sphere: the area under the $1^{st}$ peak expands and the distance in the first sphere slightly increases (Fig. 1c). Since the $1^{st}$ peak of the total distribution in the $B_2O_3$ glass corresponds only to the nearest neighbors B-O, the area under the peak is proportional to the coordination number. Thus, for the first time the coordination number can be quantitatively estimated from direct measurements. The calculated coordination number smoothly rises: the fraction of the 4-coordination boron atoms can be assessed as being about 10-15% at P~ 8 GPa T~ 300 K; about 30% at P~ 9.5 GPa, T~ 300 K, and about 45% at P~ 9.5 GPa, T~ 650 K. Consequently, the fraction of the 4-coordination boron atoms during the transformation was significantly overestimated in the previous experimental indirect measurements [4] and greatly underestimated in the previous computer simulation data [5,6]. The B-O average distance remains virtually unchanged up to 5 GPa, beginning to increase slowly (from 1.35 Å to 1.43 Å) with a further rise in pressure up to 10 GPa. This increase is also observed for $B_2O_3$ crystalline phases at the transition from a normal-pressure to high-pressure phase with the 4-coordinated boron atoms (average distances are 1.372 Å and 1.475 Å, respectively).

Glass structures under compression from *ab-initio* simulations are shown in Fig.2. The initial glass structure consists of corner-shared $BO_3$ triangles (Fig 2a). At high pressures glass structure consists mostly of $BO_4$ tetrahedra (Fig 2b,c). In Figure 1d we compare the atomic correlation function obtained in the experiment and first-principle simulations and observe reasonable agreement between them.

It should be mentioned that at the initial stage of the transformation (P ~ 5-7GPa) there is a slight shift of the B-O average distance whereas the estimated fraction of the 4-coordinated boron is close to zero. One can suppose that it is associated with the distortion of $BO_3$ triangles (boron atoms come out from the oxygen triangles planes). The *ab-initio* simulations indeed show that before transforming into the 4-fold coordinations, $BO_3$ triangles loose their planarity, as the B atom is pushed out of the plane of three O atoms starting from 4 - 5 GPa. In Figure 2d we plot the distribution of the distances between the B atom and the oxygen plane in 3-coordinated states and observe that these distances increase with pressure, reaching 0.3-0.4 Å at high pressure.

The primary qualitative change in the structure factor in the $B_2O_3$ glass with compression is observed in the ratio of the amplitudes of the $1^{st}$ and $2^{nd}$ maxima. This can be easily observed from the behavior of the experimental diffraction spectra at $2\theta = 6^0$ (see Fig.1a, Fig. 3). Fig. 3 shows the relationship between the intensities of the $1^{st}$ and $2^{nd}$ maxima vs pressure along compression-decompression cycles at room and elevated temperatures. This value incorporates the contribution from the changes occurring in both the $1^{st}$ coordination sphere and distant ones. As seen from Fig. 3, the transformation under high pressures of about 9.5 GPa has significant kinetic effects (Point A) and greatly accelerates on heating (Points B,C). The reverse structural transformation at decompression occurs in a narrower pressure range at P < 5 GPa; whereas at

P~1.5 GPa, a jump-wise behavior was observed. The structural transformation is not fully reversible, causing residual densification. The elevated temperature promotes both direct and reverse transformations (see Fig.3).

The performed in-situ high-pressure volumetric investigations essential for probing into the nature of phase transformations in the glass. They are also important for the accurate calculation of the total correlation functions shown in Fig. 1c. The corresponding data of the volumetric measurements are presented in Fig. 4. We have chosen two maxima of pressures, 9 GPa and 5.6 GPa, for two experimental runs, respectively. In this way, the experimental run I corresponds to partial coordination changes of the boron atoms in the glass, whereas the experimental run II revealed structural changes happening only in farther coordination spheres. Both experimental runs perfectly agree in the compression stages of the experiments, meaning the absolute reliability of the obtained data (especially if it is remembered that these runs were carried out in different liquid hydrostatic mediums). At pressures higher than 3 GPa, an appreciable time relaxation of density is observed in both runs. Thus, the 1 hour ageing at pressure P=5.6 GPa in run II results in the additional increase in density by 1%.

From the behavior of the "relaxed" compressibility (Fig. 4b) we may conclude that the transformation in the $B_2O_3$ glass consists of two broad overlapping transitions occurring at P > 1 GPa and P > 5 GPa. It is evident that the 1$^{st}$ transformation corresponds to the structural changes in the distant coordination spheres, while the 2$^{nd}$ one at P > 5 GPa conforms to the coordination transformation in glass. Here we note that the flat minima of the relaxed bulk modulus at P~1.7 GPa and P~6 GPa were not predicted by previous computer simulation studies [5,6]. The initial values of the relaxed bulk modulus during the onset of decompression at maximum pressures are bound to be equal to the unrelaxed values. Indeed, as observed from the Brillouin scattering data [7], the unrelaxed bulk modulus of $B_2O_3$ glass at 8-9 GPa can be estimated as 110-120 GPa, which is in perfect agreement with our experimental volumetric data ~ 110 GPa (Fig. 4b).

After decompression, the B$_2$O$_3$ glass remains densified ($\Delta\rho/\rho\sim6\%$). The residual densification relaxes at room temperature for several days. The densification is virtually the same for the different highest experimental pressures of treatment of 9 and 5.6 GPa (see Fig. 4a). Therefore, the densification is not linked to the presence of the 4-coordination boron atoms; it is related to the changes in the distant coordination spheres, such as the reduction of the proportion of the boroxol rings and the buckling of the "ribbons" formed by the B$_2$O$_3$ triangles. The coordination transition, however, is fully reversible. The distinction in the behavior of relaxed compressibility for the two experimental runs during decompression (Fig. 4b) refers to the transfer of the boron atoms from the 4- to 3-coordination state for the glass subjected to a pressure of 9 GPa. On the basis of the obtained data, it can be inferred that the reverse transformation (the change in the coordination of the boron atoms from 4 to 3) begins at P~4.5 GPa and ends at P~1 GPa. The reverse transformation is absolutely smooth in hydrostatic conditions. Therefore, the conclusion made in Ref. [7] about the first-order character of the reverse transition in the B$_2$O$_3$ glass is incompatible with our findings. The volumetric study conducted in the present work involved purely hydrostatic high-pressure conditions, whereas the Brillouin scattering study in Ref. [7] was performed under quasi-hydrostatic conditions without a pressure-transmitting medium. The abrupt changes of sound velocities, observed in Ref. [7], are thus likely to be an artifact associated with strongly non-hydrostatic conditions near the sample. In fact, the value of about 3 GPa corresponds to the strength of the gasket used in Ref. [7]. That is, at pressures of about 3 GPa during decompression, jump-like pressure variations near the sample can take place. A similar artifact, related to non-hydrostatic stresses, is jump-like changes of the structure during decompression at pressures of about 1.5 GPa, which we observed in this work (see Fig.3.) The value 1.5 GPa exactly matches the strength of the boron-epoxy gaskets used in our work.

Figure 5 presents the comparison of the experimental and calculated density and coordination changes with pressure. A reasonable agreement between experiment and first-principles calculations can be seen. Previous simulations with empirical potentials significantly underestimated the number of increased coordination, especially the results from [6], whereas previous inelastic X-ray scattering data [4] overestimated the coordination changes under compression (Fig. 5b). The somewhat different values of the density in the experiment and ab-initio simulations seen in Fig.5 are most probably due to the difference in time scales between the experiment and simulation: as mentioned above, experiments show significant time-dependent relaxation processes of both structure (see point A in Fig. 3) and density (see Fig. 4a).

Note that the recent study of the bonding changes in the $SiO_2$ glass conducted by using an inelastic x-ray scattering technique [31] similarly gives an incorrect estimate of the degree of the coordination transformation [32]. Thus, the assessments of the degree of the coordination transformations in glasses, made from inelastic x-ray scattering data [4,31,33], should be treated with caution.

The first-principles simulations enable us to analyze the behavior of glass at much larger pressures than those available in the experiments that use large volume presses. We observe that unlike its crystalline counterpart $BO_4$ tetrahedra in glass at high pressure can be connected in edges (see Figure 1b). The number of edge-shared tetrahedra increases with density, and under further compression face-shared tetrahedra with 5- and 6-coordinated B atoms appear in the glass (Fig. 1c).

Thus, the obtained in-situ structural and volumetric data give grounds to conclude that the $B_2O_3$ glass at compression experiences two pressure-overlapping, broad transformations. The first transformation corresponds to the structural changes in the distant coordination spheres, beginning from the 2[nd] one, and is only partially reversible, which leads to a residual densification in the glass after pressure treatment. The second transformation leads first to the distortion of $BO_3$ triangles, followed by a change in the B coordination. The latter transformation is fully reversible. Both direct and reverse transformations are smooth.

The behavior of $B_2O_3$ glass under pressure is in a certain sense similar to the behavior of another archetypical oxide glass, a-$SiO_2$. Under pressure, this glass equally features two overlapping diffuse transformations [34]. The 1st transformation is irreversible at room temperature without the change in the coordination of the Si - atoms, the second one is reversible; it occurs at higher pressures and is accompanied by the change in the coordination of the Si atoms from 4 to 6. The increased coordination number of cation atoms is not retained at room temperature after decompression in both the $B_2O_3$ and $SiO_2$ glasses. The behavior of these glasses greatly differs from that of their crystalline counterparts: the $B_2O_3$ II phase and stishovite $SiO_2$ can be retained at normal conditions and have high temperature stability.

In the simulations, we find permanent densification close to the experimental values; e.g. on decompression from 20 GPa, we observed 8% densification. If decompression is simulated at low 10 K temperature, the permanent densification increases, but only slightly (to 12%), implying that high-coordinated glass is almost not quenchable.

In our ab-initio simulations at very high pressure --in the 150-250 GPa range-- we find 5- and 6-fold coordinated B atoms as well as 4-fold coordinated O atoms. The appearance of these states can stimulate the search for the new glassy and crystalline forms of $B_2O_3$ with high coordination under megabar pressures. Such phases may possibly be quenchable, and should possess very high hardness and elastic moduli.

METHODS

Heat treatment of the $B_2O_3$ glass at T~ 600-700 K within 20-40 hours or at T~ 900 K within 5-10 hours has made it possible to obtain dehydrated specimens which were rapidly (~1 min) mounted into a high pressure cell. For the structural study, the glassy samples were 2 mm dia, 1.2 mm high. A cubic press was used for generating high pressure up to 10 GPa in amorphous boron-epoxy cubes with graphite heaters. The temperature was measured by a chromel-alumel thermocouple. The pressure was determined from the equation of state of NaCl. The in-situ structural investigations of the $B_2O_3$ glass were carried out by the energy-dispersive x-ray diffraction method in the SMAP180 press at the SPring-8 synchrotron radiation facility at the BL14B1 beamline. The spectra were registered at 10 different diffraction angles ($2\theta = 3^0, 4^0, 5^0, 6^0, 8^0, 10^0, 12^0, 16^0, 20^0, 24^0$), which made possible the restoration of the glass structural factor and respective total distribution function with high accuracy. The X-ray diffraction spectra were registered at 15 pressure points, both during compression and decompression. A few independent sets of experiments at different temperatures were carried out.

The in-situ volumetric measurements of the $B_2O_3$ glass were conducted by using the toroid high-pressure apparatus [25], generating up to 9 GPa pressure at room temperature. The strain gauge technique [26] was used to measure the sample size. The sample of the $B_2O_3$ glass of about 3 mm in size was covered with special protective lacquer and placed into a pure hydrostatic high-pressure medium (ethanol-methanol mixture at pressures of up to 9 GPa, pentan-isopentan mixture at pressures up to 5.6 GPa). The absolute accuracy of the relative volume measurements was 0.1%, the relative accuracy (measurement sensitivity) was about $10^{-3}$%.

In the simulation, we used empirical interatomic potentials [27] to equilibrate a 135-atom liquid at 5000 K for 1 ns. We then used the *ab-initio* simulation to cool down the liquid to obtain glass. We have used the SIESTA method [28], an implementation of density functional theory [29], and employed the general gradient approximation to the exchange-correlation energy. The Kohn-Sham eigenstates were expanded in a localized basis set of numerical orbitals. The DZP-basis set was optimized by minimizing the energy of a boroxol molecule capped by hydrogens. Quenching the liquid to 300 K and relaxing the resulting structure produced glass with density of about 1.9 g/cm3, in a reasonably good agreement with the experimental value of 1.8 g/cm3. In terms of the number of boroxol rings, the final glass structure was similar to that in previous *ab-initio* study [30]. This structure was used in high-pressure *ab-initio* simulations at 300 K. Each pressure point was simulated for 3-4 ps on Cambridge HPC using 32 parallel processors. The total computer time was equivalent to about 3 years of simulation on a single processor.

Acknowledgments:

The authors wish to thank T. Hattori, S.V. Popova, and W. Utsumi for their help and to P. McMillan and M. Wilding for valuable discussions. The synchrotron radiation experiments were performed at the SPring-8 with the approval of the JAEA (project 2006B-E18). The work has been supported by the RFBR (05-02-16596 and 07-02-01275), by the Programs of the Presidium of RAS, by EPSRC and by UK Royal Society.

**Figures**

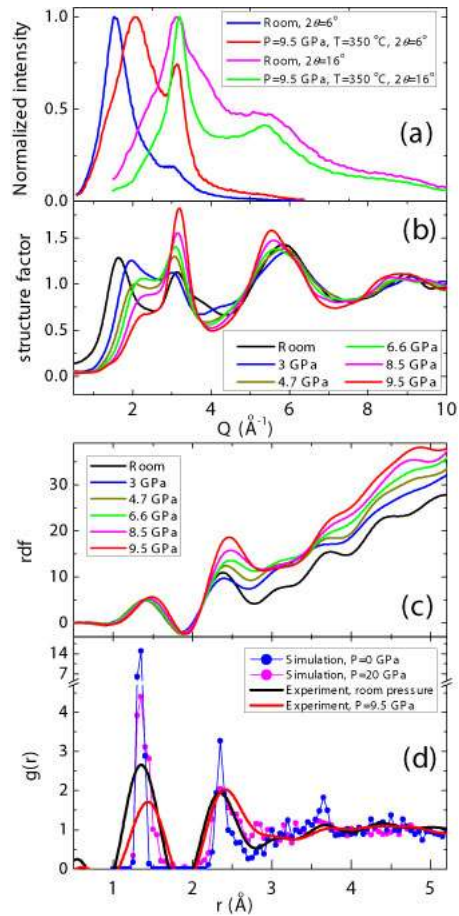

Fig. 1. Examples of the energy dispersive x-ray diffraction (EDXD) data for the glassy $B_2O_3$ (a), measured at different pressures at the two angles of the detector; calculated from the experimental EDXD data structure factor (b) and total radial distribution function (c), and comparison of the experimental and computer simulated total correlation functions for different pressures (d). All experimental curves for $P$=9.5 GPa correspond to the sample obtained after heating to 350 ºC.

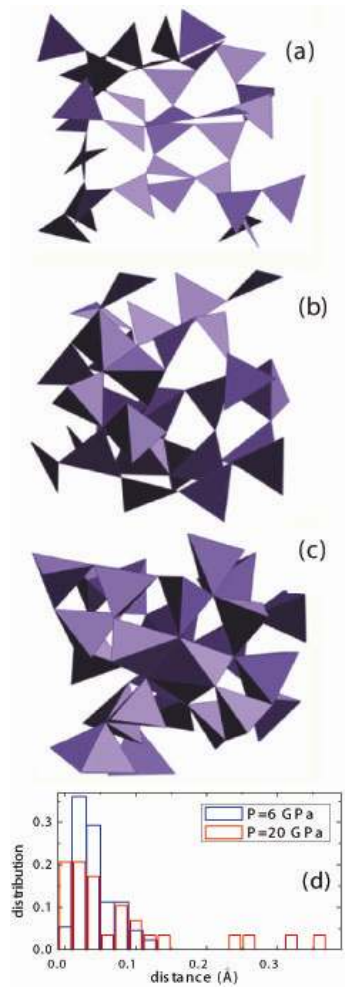

Fig. 2. Polyhedron representation of the simulated $B_2O_3$ glass structure at zero pressure (a), $P$=20 GPa (b), and $P$=200 GPa (c), where triangles correspond to the $BO_3$ structure units (3-fold coordinated boron), tetrahedrons - to $BO_4$, and there are the single $BO_5$ and $BO_6$ units on (c). The plot (d) shows the distribution of distances from B atom to the $O_3$ plane in the $BO_3$ units for 2 different pressures.

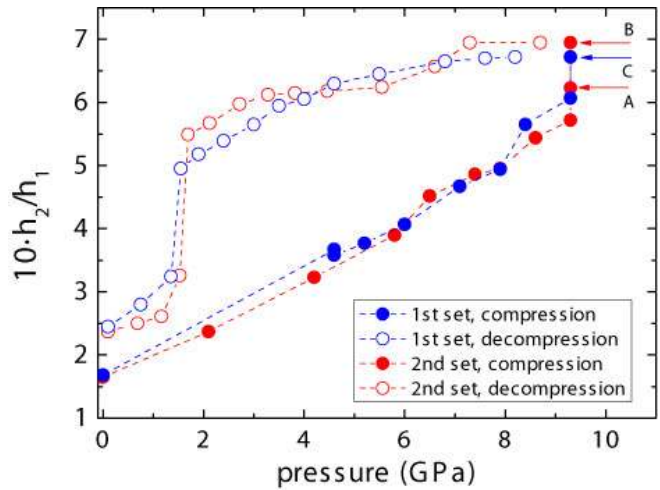

Fig. 3. Pressure dependence of the ratio of amplitudes of the 2$^{nd}$ and 1$^{st}$ diffraction peaks in the EDXD data measured at the angle of the detector $2\theta=6°$ [see Fig. 1 (a)] for the two independent compression-decompression runs at room temperature (red points) and $T=200$ °C (blue points). Point A corresponds to the sample obtained after 3 hours relaxation at the fixed pressure ($P=9.5$ GPa), points B and C correspond to the sample states after heating to 350 °C under pressure.

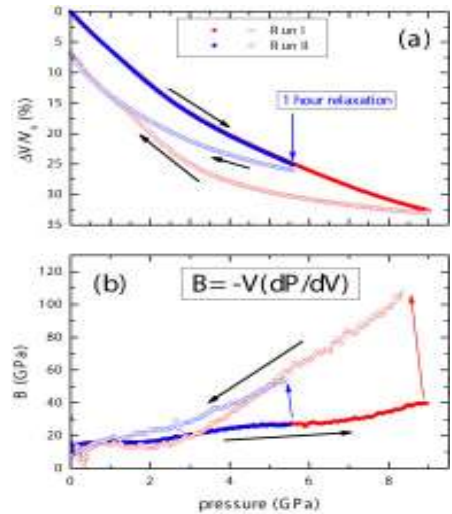

Fig. 4. Results of the *in situ* volumetric measurements of the glassy B$_2$O$_3$ under pressure (a) in the two different runs of compression (solid symbols) and decompression (open symbols), and pressure dependences of the bulk modulus (b) obtained by the direct numerical differentiation of the curves from the plot (a) (relaxed modulus). The significant jumps of the effective bulk modulus between the final of compression and onset of decompression for both runs correspond to the jumps between relaxed and almost unrelaxed values (see the text).

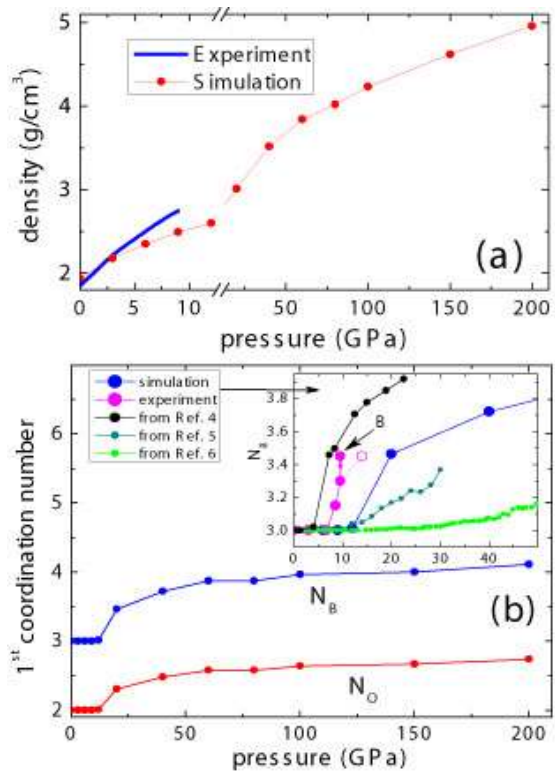

Fig. 5. Comparison of pressure dependences of the experimental (Fig. 4, compression in the run I) and simulated density (a), and simulated pressure dependences of the averaged 1st coordination numbers for B and O (b). The inset in (b) shows pressure dependences of the 1st coordination number for B from the current experiment (x-ray diffraction) and simulation and according to the data from Refs. [4-6]. Point B in the experimental dependence corresponds to that equally marked in Fig. 3 (i.e., after heating), whereas the open point is the estimation of the point B position which should be observed under room-temperature compression.